# Can the interface between a non-ideal ferromagnet and a semiconductor quantum wire act as an ideal spin filter?

Marc Cahay[#] and Supriyo Bandyopadhyay[*]

[#]Department of Electrical and Computer Engineering and Computer Science
University of Cincinnati, Cincinnati, OH 45221, USA, E-mail: marc.cahay@uc.edu
[*]Department of Electrical and Computer Engineering
Virginia Commonwealth University, Richmond VA 23284, USA, E-mail: sbandy@vcu.edu

*Abstract* -- The problem of spin injection across the interface between a non-ideal ferromagnet and a semiconductor (paramagnetic) quantum wire is examined in the presence of Rashba spin orbit interaction in the wire and an axial magnetic field along the direction of current flow. This magnetic field is caused by the ferromagnet magnetized along the wire axis. At low temperatures and for certain injection energies, the interface can act as an ideal spin filter allowing injection *only* from the majority spin band of the ferromagnet. Thus, 100% spin filtering can take place even though the ferromagnet itself is less than 100% spin polarized. Below a critical value of the axial magnetic field, there are two injection energies for which ideal (100%) spin filtering is possible; above this critical field there is only one such injection energy.

## I. Introduction

One of the most important challenges in spintronics is the ability to selectively extract spin of a particular polarization from a ferromagnet and inject it in a semiconducting paramagnet. This problem has received increasing attention over the last five years since it is critical to the implementation of many spintronic devices. Several recent experimental investigations have shown successful spin injection into semiconductors from ferromagnets, with injection efficiencies as high as 90% for semiconducting ferromagnets and 32% for metallic ferromagnets coupled with a Schottky or tunnel barrier [1]. Several theoretical models have also been developed to elucidate spin injection across specular and disordered ferromagnet-semiconductor (Fe/Sm) interfaces, some of which have been based on a simple Stoner model of the ferromagnetic contact [2,3] while others have included the full electronic band structure of the contact [4,5].

In this paper, we study the following problem. Is it possible to perform ideal spin filtering across the interface between a non-ideal ferromagnet and a paramagnet? That is, can we selectively inject electrons only from the majority spin band of the non-ideal ferromagnet which is not a half metal and does not itself possess 100% spin polarization? This problem is of relevance to many proposed spintronic devices such as polarized light emitting diodes and spin analogs of electro-optic modulators [6], which operate best when the non-ideal ferromagnet acts as an ideal spin polarizing injector that injects electrons from the majority band *only* while blocking those from the minority spin band. Here we show that such ideal spin filtering is possible for certain injection energies at very low temperature.

## II. THEORY

The system we investigate is shown in Fig. 1. It consists of a semiconductor quantum wire interfaced with a ferromagnetic contact. There is a Rashba interaction [7] in the wire caused by a structural inversion asymmetry arising from a transverse electrostatic field (in the y-direction). This field could be due to either the natural asymmetry of the confining potential (triangular well) or could be externally imposed with a gate terminal.

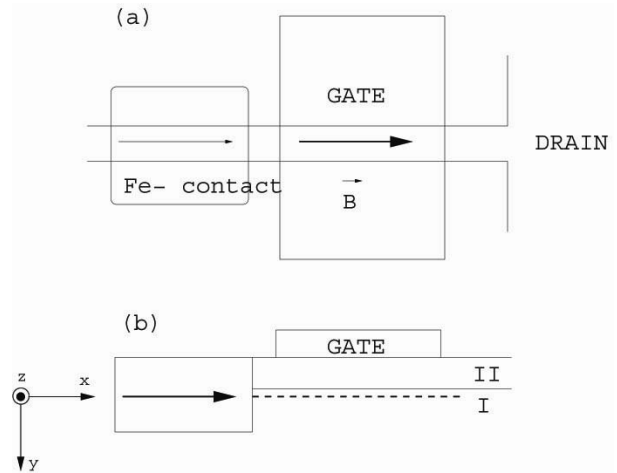

*Fig. 1: Ferromagnetic/Semiconducting Quantum Wire Contact. (a) Top view of the structure showing the ferromagnetic contact with its magnetization in the direction of current flow in the quantum wire. (b) Cross-sectional view of the structure showing as thick dashed line the quasi one-dimensional electron gas in the wire.*

In ref. [8,9], we derived the energy dispersion relations (E-k) for the lowest energy bands in the semiconductor quantum wire. The bands have the general shape shown in Figure 2. The lowest subband has *two* minima at wavevectors

$$k_{min} = \pm k_R \sqrt{[1 - (g\mu_B B/16\delta_R)]}$$

where $g$ = Landé g-factor, $\mu_B$ = Bohr magnetron, $B$ is the flux density associated with the axial magnetic field, $\delta_R = \hbar^2 k_R^2/2m^*$, $k_R = m^*\alpha_R/\hbar^2$, and $\alpha_R$ is the strength of the Rashba spin orbit interaction [10]. The presence of two minima gives rise to a "camel-back" shape of the lower subband as long as the magnetic field strength is below a critical value $B_c$. Above the critical strength, the camel-back feature disappears, as shown in Fig. 2 (bottom). It is easy to see that $B_c = 16\delta_R/(g\mu_B B)$ since when $B > B_c$, $k_{min}$ becomes imaginary.

There are three critical energies shown in Fig. 2. The energy at the bottom of the higher subband is $E_3$, the energy at the camel-back is $E_2$ and the energy of the double minima in the lower subband is $E_1$. Using the dispersion relations derived in ref. [8], we can show that these three energies are: $E_1 = \hbar\omega/2 + \Delta E_c - [\delta_R + (g\mu_B B)^2/16\delta_R]$, $E_2 = \hbar\omega/2 + \Delta E_c - g\mu_B B/2$ and $E_3 = \hbar\omega/2 + \Delta E_c + g\mu_B B/2$, where $\hbar\omega$ is the energy separation between the subbands in the quantum wire in the absence of any magnetic field and spin-orbit interaction, and $\Delta E_c$ is the energy separation between the bulk conduction band edge in the semiconductor and the bottom of the majority spin band in the ferromagnet as shown in Fig. 2.

Consider the situation when $B < B_c$ (Fig. 2; top). If the Fermi level $E_f$ is above $E_3$, there will be two (spin-resolved) current-carrying channels in the quantum wire since $E_f$ intersects the two E-k curves at two positive values of the wavevector. In the range $[E_2, E_3]$, there is only one current-carrying channel since the upper band becomes evanescent with an imaginary wavevector that becomes exactly zero when $E_f = E_2$ or $E_3$.

In the energy range $[E_1, E_2]$, we recover two current-carrying channels since the Fermi level intersects the energy bands at two points that correspond to positive current. Below $E_1$, there are no current-carrying states, and the conductance drops sharply. When $B > B_c$ (Fig. 2; bottom), there are two current-carrying channels when $E_f > E_3$ and only one otherwise.

We now set forth to calculate the interface conductances for electrons incident from the majority and the minority spin bands in the ferromagnet. This requires solving the tunneling problem across the Fe/Sm interface. We model the ferromagnetic contact by the Stoner-Wohlfarth model. The magnetization of the contact is assumed to be along the wire axis (x-direction) so that the majority carriers are +x-polarized electrons and minority carriers are -x-polarized. Their bands are offset by an exchange splitting energy $\Delta$, shown in Fig. 2.

In the semiconductor quantum wire, when $B > B_c$, the x-component of the wavefunction of an electron, having an energy $E_f$ in the contact, at a position $x$ along the the channel is given by [8]

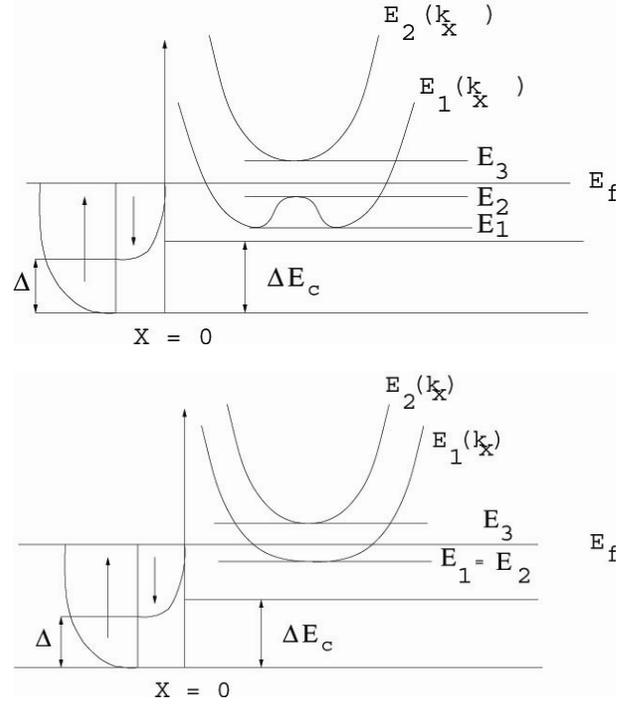

Fig. 2: Energy band diagram across the ferromagnetic contact/semiconducting channel. $\Delta$ is the exchange splitting energy in the ferromagnetic contact and $\Delta E_c$ is the energy difference between the band bottoms in the semiconductor and ferromagnetic contact. Also shown are the energy dispersion relationships on both sides of the Fm/Sm interface for $B < B_c$ (top figure) and $> B_c$ (bottom figure).

$$\varphi(x) = T_1 \begin{bmatrix} C_1(k_{x,1}) \\ C_1^{'}(k_{x,1}) \end{bmatrix} e^{ik_{x,1}x} + T_2 \begin{bmatrix} C_1(k_{x,1}) \\ C_1^{'}(k_{x,1}) \end{bmatrix} e^{ik_{x,2}x}$$

where $k_{x,1}$ and $k_{x,2}$ are the wavevectors in the lower and upper subbands where the Fermi level $E_f$ intersects the E-k dispersion relationships shown in Fig. 2; $T_1$ and $T_2$ are the corresponding transmission amplitudes into these states and the spinors $[C_1(k_{x,1}), C_1^{'}(k_{x,1})]^T$, $[C_2(k_{x,2}), C_2^{'}(k_{x,2})]^T$ (where $T$ stands for transpose) are the eigenspinors for these states. Expressions for these quantities can be found in ref. [8]. Note that $k_{x,2}$ becomes imaginary when $E_f < E_3$ since the upper band becomes evanescent.

When $B < B_c$, the above equation is valid as long as $E_f > E_2$. For $E_f < E_2$, the above equation should be replaced with

$$\varphi(x) = T_I \begin{bmatrix} C_1(k_{x,1}) \\ C_1^{'}(k_{x,1}) \end{bmatrix} e^{ik_{x,1}x} + T_{II} \begin{bmatrix} C_1(-k_{x,1}) \\ C_1^{'}(-k_{x,1}) \end{bmatrix} e^{-ik_{x,2}x}$$

where $k_{x,1}$ and $k_{x,2}$ are the magnitudes of the wavevectors in the lower subband where $E_f$ intersects the E-k curve shown in Fig. 2, and $T_I$, $T_{II}$ are the transmission amplitudes into these two states. The selection of the minus sign in front of $k_{x,2}$ in the second part of the wavefunction given above is required to ensure positive group velocity of the second propagating spin subband in the energy range [$E_1$, $E_2$].

In the ferromagnet, for an incident +x-polarized electron, the x-component of the wavefunction is given by

$$\varphi_{+x}(x) = \frac{1}{\sqrt{2}} \begin{bmatrix} 1 \\ 1 \end{bmatrix} e^{ik_{x,u}x} + \frac{R_1}{\sqrt{2}} \begin{bmatrix} 1 \\ 1 \end{bmatrix} e^{-ik_{x,u}x} + \frac{R_2}{\sqrt{2}} \begin{bmatrix} 1 \\ -1 \end{bmatrix} e^{-ik_{x,d}x}$$

where $R_1$ is the reflection amplitude into the +x-polarized band and $R_2$ is the reflection amplitude in the -x-polarized band.

For the incident -x-polarized electron, the wavefunction in the ferromagnet is given by

$$\varphi_{-x}(x) = \frac{1}{\sqrt{2}} \begin{bmatrix} 1 \\ -1 \end{bmatrix} e^{ik_{x,d}x} + \frac{R'_2}{\sqrt{2}} \begin{bmatrix} 1 \\ -1 \end{bmatrix} e^{-ik_{x,d}x} + \frac{R'_1}{\sqrt{2}} \begin{bmatrix} 1 \\ 1 \end{bmatrix} e^{-ik_{x,u}x}$$

where $R'_1$ is the reflection amplitude into the +x-polarized band and $R'_2$ is the reflection amplitude in the -x-polarized band.

The wavevectors

$$k_{x,u} = \frac{1}{\hbar}\sqrt{2m_0 E_f} \quad ; \quad k_{x,d} = \frac{1}{\hbar}\sqrt{2m_0(E_f - \Delta)}$$

are the x components of the wavevectors in the +x (majority spin) and -x-polarized (minority spin) energy bands in the ferromagnet, respectively.

The sets of four unknowns (reflection and transmission amplitudes) are found by enforcing continuity of the wavefunction and the current density across the Fe/Sm interface which results in a system of 4 x 4 system of coupled equations [8].

At T = 0 K, the (linear response) interface conductance for electrons incident from either the +x-polarized band or the -x-polarized band is found from the Landauer formula. For each band, the transmission coefficient is the ratio of the current amplitude in the semiconducting channel divided by the current amplitude of the incident beam in the ferromagnetic contact. The current expression calculated in the semiconductor is quite complicated [11] since the two current-carrying channels in the semiconductor are not orthogonal to each other. However, the two current carrying modes (majority and minority spin bands) in the ferromagnet are orthogonal. Therefore, it is easier to calculate the conductance across the interface, for majority and minority incident spins, as follows:

$$G_{+x} = \frac{e^2}{h}\left(1 - |R_1|^2 - \frac{k_{x,d}}{k_{x,u}}|R_2|^2\right)$$

$$G_{-x} = \frac{e^2}{h}\left(1 - |R'_2|^2 - \frac{k_{x,u}}{k_{x,d}}|R'_1|^2\right)$$

The spin filtering efficiency η is defined as

$$\eta = (G_{+x} - G_{-x})/(G_{+x} - G_{-x})$$

In order to calculate η, we use parameters relevant to an interface between an Fe contact and an InAs quantum wire as shown in Table 1. Sources for the various values are cited in the bibliography. With these values, the spin polarization in the ferromagnet is 86%. We also include the effect of an interface potential barrier, as was done in ref. [8,12], by modeling it as a delta-function potential $V_I(x) = V_0\delta(x)$. This barrier has a major effect on η. Following usual practice, we parameterize this barrier with a dimensionless quantity Z defined as $Z = 2m_f^* V_0/\hbar^2$ where $m_f^*$ is the effective mass of electrons in the ferromagnet. Typical values of Z range from 0 to 2 [12].

TABLE 1

PARAMETERS FOR FE/INAS INTERFACE

| | |
|---|---|
| Fermi energy $E_f$ (eV) | 4.2 |
| Exchange splitting Δ (eV) | 3.46 |
| Rashba energy $\delta_R$ (eV) | 0.2 |
| Landé g-factor | 15 |
| Effective mass in Fe ($m_0$) | 1.0 |
| Effective mass in InAs ($m_0$) | 0.024 |
| Subband separation $\hbar\omega$ (meV) | 10.0 |

With the parameters in Table 1, we calculate η as a function of $\Delta E_c$ and the results are shown in Fig. 3.

Varying $\Delta E_c$ results in sweeping the Fermi level through the energy dispersion relations shown in Fig. 2. We repeat the calculation for two different values of the magnetic field, above and below the critical field. For simplicity, we assume that the Rashba energy $\delta_R$ is independent of the gate potential. This is a realistic assumption since the Rashba interaction typically has a very weak dependence on gate potential [13]. In principle, the interface barrier $V_0$ (hence the parameter Z) can also have a weak dependence on gate potential, but we neglect that here. For the parameters in Table 1, the critical magnetic field $B_c = 0.92$ Tesla.

When $B < B_c$, η reaches 100%, *independent* of the interface potential barrier (or Z), at values of $\Delta E_c$ corresponding to the situations when the Fermi level $E_f$ coincides with either $E_1$ or $E_3$. Moreover, η falls to a minimum when $E_f$ is aligned with $E_2$. For $B > B_c$, η reaches a maximum when $E_f$ coincides with $E_3$. We also observe that the spin injection efficiency is larger for larger value of Z, i.e, the presence of a tunneling barrier at the interface helps increase the injection efficiency, as first pointed out by Rashba [14].

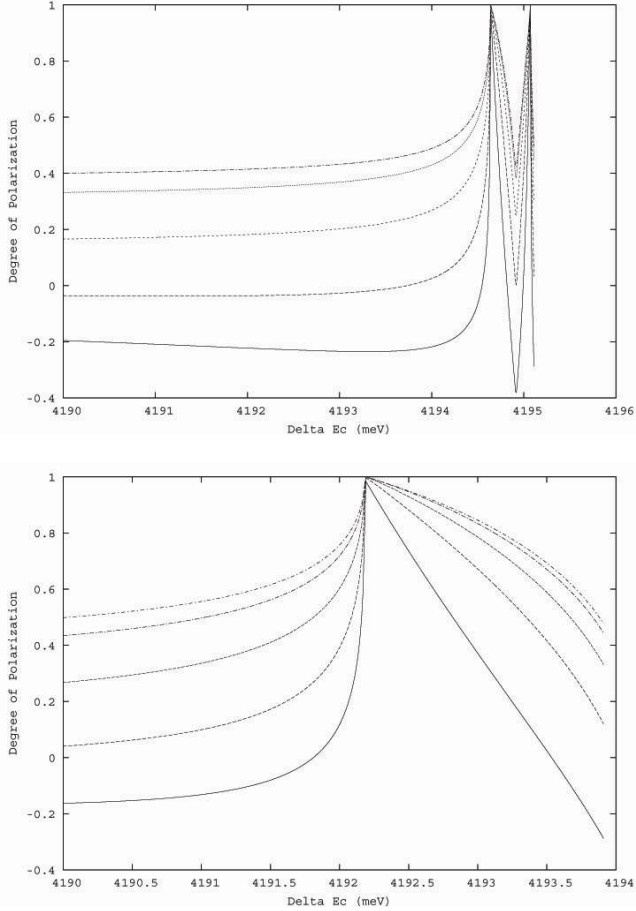

*Fig. 3: Spin filtering efficiency η as a function of ΔE$_c$ and normalized interface potential barrier Z: (top) for a magnetic field B = 0.5 Tesla which is below the critical B$_c$, and (bottom) for a magnetic field B = 2 Tesla which is above the critical value B$_c$. The calculations are for absolute zero temperature. From bottom to top, Z = 0.0, 0.25, 0.5, 1.0, and 2.0, respectively.*

The reason for the 100% maximum efficiency is that, for electrons incident with either energy E$_2$ or E$_3$ (Zeeman energies), the spinor $[C_1(k_{x,1}), C_1'(k_{x,1})]^T$ reduces to $(1/\sqrt{2})[1,-1]^T$ (see Equation (8) in ref. 8]. On the other hand, $[C_2(k_{x,2}), C_2'(k_{x,2})]^T$ is not equal to $(1/\sqrt{2})[1,-1]^T$ for E$_f$ = E$_2$ or E$_3$. Hence, the continuity of the wavefunction at x = 0 for an electron incident in the minority spin-band (-x-polarized) requires that not only T$_{II}$ must be zero, but T$_I$ must also be zero otherwise the wavefunction describing the electron in the semiconducting quantum wire cannot be normalized. As a result, G$_{-x}$ must be identically zero for E$_f$ = E$_2$ or E$_3$. Hence, η is equal to 100% at those energies.

## III. CONCLUSION

In conclusion, we have shown that it is possible to achieve ~100% spin filtering efficiency at low temperatures with a non-ideal ferromagnet at certain critical injection energies. The injection energy can be tuned with a back-gate potential. To our knowledge, no spin injection experiment has employed this technique to optimize spin filtering. This surprising and unexpected result assumes practical importance in view of the fact that ideal half-metallic ferromagnets with 100% spin polarization are rare and difficult to integrate with semiconductors. Therefore, if non-ideal ferromagnets with much less than 100% spin polarization can result in nearly 100% spin filtering, then that is a major technological feat.

REFERENCES

1) J. Seufert, et al., "Spin injection into a single self-assembled quantum dot," *Phys. Rev. B*, Vol. 69, 035311, January 2004 and references therein.
2) C.-M. Hu and T. Matsuyama, "Spin Injection Across a Heterojunction: A Ballistic Picture," *Phys. Rev. Lett.*, Vol. 87, no. 6, 066803, August 2001.
3) G. Schmidt and L. W. Molenkamp, "Spin injection into semiconductors using dilute magnetic semiconductors," *Semicond. Sci. & Tech.*, Vol. 17, no. 4, pp. 310-321, April 2002.
4) M. Zwierzycki, et al., "Spin-injection through an Fe/InAs interface," *Phys. Rev. B*, Vol. 67, no. 9, 092401, March 2003.
5) O. Wunnicke, et al., "Ballistic spin injection from Fe(001) into ZnSe and GaAs," *Phys. Rev. B*, Vol. 65, no. 24, 241306(R), June 2002.
6) S. Datta and B. Das, "Electronic Analog of the Electro-optic Modulator," *Appl. Phys. Lett.*, Vol. 56, no. 7, pp. 665-667, February 1990.
7) Y. A. Bychkov and E. I. Rashba, "Oscillatory effects and the magnetic susceptibility of carriers in inversion layers," *J. Phys. C*, Vol.17, no. 33, pp. 6039-6045, November 1984.
8) M. Cahay and S. Bandyopadhyay, "Phase coherent quantum mechanical spin transport in a weakly disordered quasi one-dimensional channel," *Phys. Rev. B*, Vol. 69, no. 4, 045303, January 2004.
9) M. Cahay and S. Bandyopadhyay, "Conductance modulation of spin interferometers," *Phys. Rev. B*, Vol. 68, no. 11, 115316, September 2003.
10) The expression for k$_{min}$ given in Ref. 8 is incorrect. The expression in this paper is correct.
11) M. Cahay and S. Bandyopadhyay (unpublished).
12) Th. Sch\"apers, et al., "Interference ferromagnet semiconductor ferromagnet spin field-effect transistor," *Phys. Rev. B*, Vol. 64, no. 12, 125314, September 2001.
13) J. Nitta, T. Akazaki, H. Takayanagi, and T. Enoki, "Gate control of spin-orbit interaction in an inverted In$_{0.53}$Ga$_{0.47}$As/In$_{0.52}$Al$_{0.48}$As heterostructure," *Phys. Rev. Lett.*, Vol. 78, no. 7, pp.1335-1338, February 1997.
14) E.I. Rashba, "Theory of electrical spin injection: tunnel contacts as a solution of the conductivity mismatch problem," *Phys. Rev. B*, Vol.68, R16267-R16270, December 2000.